\documentstyle[sprocl]{article}
\def\ga{\gamma}
\def\de{\delta}
\def\ep{\epsilon}

\def\la{\lambda}

\def\ph{\phi}

\def\ch{\chi}
\def\ps{\psi}

\def\Ga{\Gamma}
\def\De{\Delta}

\def\La{\Lambda}

\def\cl{{\cal L}}
\def\fr#1#2{{{#1} \over {#2}}}
\def\prt{\partial}

\def\vev#1{\langle {#1}\rangle}

\def\frac#1#2{{\textstyle{{#1}\over {#2}}}}

\def\lsim{\mathrel{\rlap{\lower4pt\hbox{\hskip1pt$\sim$}}
    \raise1pt\hbox{$<$}}}
\def\gsim{\mathrel{\rlap{\lower4pt\hbox{\hskip1pt$\sim$}}
    \raise1pt\hbox{$>$}}}
\def\sqr#1#2{{\vcenter{\vbox{\hrule height.#2pt
         \hbox{\vrule width.#2pt height#1pt \kern#1pt
         \vrule width.#2pt}
         \hrule height.#2pt}}}}

\def\Re{\hbox{Re}\,}
\def\Im{\hbox{Im}\,}

\newcommand{\beq}{\begin{equation}}
\newcommand{\eeq}{\end{equation}}
\newcommand{\bea}{\begin{eqnarray}}
\newcommand{\eea}{\end{eqnarray}}
\newcommand{\rf}[1]{(\ref{#1})}

\begin{document}

\begin{flushright}
IUHET 365 \\
April 1997
\end{flushright}
\bigskip

\title{SOME QUESTIONS AND ANSWERS ABOUT CPT VIOLATION}
\author{V.A. KOSTELECK\'Y}
\address{Physics Department, Indiana University,\\
Bloomington, IN 47405, U.S.A.}

\maketitle\abstracts{
Minuscule violations of CPT and Lorentz invariance 
might arise in an extension of the standard model 
as suppressed effects from a more fundamental theory.
In this contribution to the CarruthersFest,
I present and answer some questions 
about CPT and the possibility of its violation.}

\section{Introduction}

It is a pleasure for me to join in celebrating Pete's
61st birthday. 
In the spirit of Pete's approach to physics and to life,
here are answers to some of the questions you always
had about CPT and its violation 
(but were afraid to ask).

\section{Basics}

\noindent \it
What is the CPT theorem?
\rm
\\
The CPT theorem states 
that the product CPT of three discrete transformations,
namely, charge conjugation C, 
parity reflection P, 
and time reversal T,
is an exact symmetry of local relativistic field theories 
of point particles.\cite{cpt}
The theorem has withstood numerous 
high-precision experimental tests.\cite{pdg}
Indeed,
CPT remains to date the only combination of C, P, T
that is observed as an exact symmetry of nature.

\medskip\noindent \it
Why consider CPT violation if the theorem says it isn't broken?
\rm
\\
The CPT theorem is a general result holding for 
relativistic particle theories,
and it can be experimentally tested to great accuracy.
These facts make CPT violation an excellent candidate
signature for unconventional physics,
such as might arise in a fundamental theory
based on extended objects like, say, 
strings.\cite{kp1}$^{\!-\,}$\cite{kp3}

\medskip\noindent \it
How could CPT be violated?
\rm
\\
Theories disobeying any of the assumptions
that enter the CPT theorem
could violate CPT.
In fact,
one of the earliest explicit examples of CPT violation
was presented by Pete.\cite{pete}
It is a class of particle theories that naively appear normal
but that entail nonlocal interactions,
excluded by assumption in the CPT theorem.

It is also possible to produce theories violating CPT
for relatively subtle technical reasons.
For example,
the CPT theorem assumes that fields appear
in finite-dimensional representations of the Lorentz group.
Certain theories involving infinite-dimensional representations
break CPT.\cite{ot}
Also,
since the CPT theorem holds within quantum field theory,
it may be theoretically feasible to violate CPT 
if conventional quantum mechanics fails.
This possibility has been suggested 
in the context of quantum gravity.\cite{sh}

\medskip\noindent \it
Are there cases where CPT violation might be 
physically interesting?
\rm
\\
A somewhat unexpected example where the CPT theorem
does not appear to apply directly 
is ordinary quantum chromodynamics
(QCD).
The CPT theorem makes assumptions about 
the correspondence between the asymptotic Hilbert space 
and the fields in the theory
that are open to question in QCD because of confinement.
Rob Potting and I thought about this issue back in 1990 
during our investigations of CPT symmetry in strings,
but we didn't find an explicit proof 
or refutation of CPT invariance in QCD.
To my knowledge, 
this aspect of CPT in QCD 
(and other confining theories)
remains an open issue at present.

A particularly interesting and conceivably physical situation 
is spontaneous CPT violation.\cite{kp1,kp2} 
In this case
the dynamics of the action remains CPT invariant,
which means many desirable properties of the theory
are preserved.
The violation occurs spontaneously in the solutions
of the equations of motion,
like the spontaneous breaking of the electroweak
gauge group in the standard model.
This type of CPT violation is a possibility in string theory,
where the usual axioms of the CPT theorem
may be modified because strings are extended objects.

\section{Spontaneous CPT Violation}

\noindent \it
How could spontaneous CPT violation occur?
\rm
\\
Suppose a higher-dimensional action
that is Lorentz- and CPT-invariant 
underlies nature.
The higher-dimensional Lorentz group
would presumably be spontaneously broken
by the solution to this theory,
since it must represent our apparently four-dimensional world.
This may induce spontaneous CPT breaking.

As an example,
strings naturally exist in higher dimensions.
Spontaneous Lorentz violation\cite{ks}
is possible in string theory
because string interactions exist
that can trigger nonzero expectation values 
for Lorentz-tensor fields.
Comparable interactions don't appear
in conventional four-dimensional 
renormalizable gauge theories.
If one or more of these tensors
has an odd number of spacetime indices,
CPT is also spontaneously broken.\cite{kp1,kp2}

\medskip\noindent \it
Can these ideas be verified explicitly in string theory?
\rm
\\
For the field theory of the open bosonic string,
the explicit action and equations of motion
can be derived analytically  
for particle fields below some fixed level number $N$.
Solutions have been found and
compared for different $N$,
in some circumstances to a depth of over 20,000
terms in the static potential.\cite{ks2,kp4}
These solutions include ones spontaneously breaking
Lorentz and CPT invariance
that persist as $N$ is increased.

\medskip\noindent \it
Does spontaneous CPT violation have to come from a string theory?
\rm
\\
If spontaneous CPT violation emerges within 
a higher-dimensional theory
then a string origin would presently seem 
to be the only possibility,
since to my knowledge 
no other consistent candidate theories exist.

\medskip\noindent \it
Does spontaneous CPT violation 
imply Lorentz violation or vice versa?
\rm
\\
If the spontaneous CPT breaking arises from
nonzero expectations of Lorentz tensors,
then Lorentz invariance is necessarily 
spontaneously violated too.
However,
the converse is false,
because expectation values of Lorentz tensors
with an even number of indices preserve CPT.

\medskip\noindent \it
Does spontaneous Lorentz violation imply causality 
is destroyed?
\rm
\\
To my knowledge,
there are no theoretical (or experimental) reasons
to exclude (small)
\it spontaneous \rm Lorentz violation.
Unlike other kinds of Lorentz breaking
that do violence to accepted notions,
spontaneous breaking is merely a feature of the
solutions to the theory.
The underlying dynamics remains Lorentz invariant.
Indeed,
it is possible to verify explicitly
that microcausality is preserved
in certain simple models arising
from spontaneous Lorentz breaking.\cite{cksm}

Other considerations make it seem very unlikely
that a fundamental problem exists
with the notion of spontaneous Lorentz violation.
For example,
the physics of a particle moving inside a biaxial crystal
need not be (rotation or boost) Lorentz covariant,
but this is merely a reflection of the presence 
of the background crystal fields
and does not affect causality.
Nonzero Lorentz-tensor expectation values
throughout spacetime are similar in some respects 
and therefore also might be expected to have benign effects.

\medskip\noindent \it
If spontaneous CPT/Lorentz breaking occurs,
where are the Goldstone bosons?
\rm
\\
Goldstone's theorem does not apply to discrete symmetries
like CPT.
If Lorentz invariance is treated as a global symmetry,
its spontaneous breaking would indeed
produce massless excitations,
carrying quantum numbers related to the graviton.
However,
Lorentz invariance is believed to be local.
In vector gauge theories,
Goldstone bosons would be absorbed by the gauge fields,
which become massive
through the Higgs mechanism.
In the present case,
the graviton propagator is affected
but no graviton mass is generated.\cite{ks}

\section{Standard-Model Extension}

\noindent \it
If spontaneous CPT breaking occurs in higher dimensions,
is it observable?
\rm
\\
If the mechanism of spontaneous Lorentz and CPT violation
occurs in a higher-dimensional theory,
it would seem likely to involve 
the four physical dimensions too.
However, 
as neither Lorentz nor CPT breaking 
have been experimentally observed,
any effects at the level of the standard model 
must be highly suppressed.\cite{kp1}$^{\!-\,}$\cite{kp3}

\medskip\noindent \it
What would be the scale of the suppression?
\rm
\\
Taking the scale governing the fundamental theory
as the Planck mass $m_{\rm Pl}$
and denoting the electroweak scale by $m_{\rm ew}$,
the natural suppression factor 
for Planck-scale effects in the standard model 
is\cite{kp1}$^{\!-\,}$\cite{kp3}
$m_{\rm ew}/m_{\rm Pl} \simeq 10^{-17}$.
A factor this small means 
that only a few Lorentz and CPT-violating effects 
are likely to be observable.

\medskip\noindent \it
How would effects in the fundamental theory
appear in a low-energy theory?
\rm
\\
The fermionic sector of the 
four-dimensional low-energy effective theory
might,
for example,
contain terms of the form\cite{kp2,kp3}
\beq
\cl \sim \fr {\la} {M^k} 
\vev{T}\cdot\overline{\ps}\Ga(i\prt )^k\ch
+ {\textstyle h.c.}
\quad .
\label{a}
\eeq
Here, a fermion bilinear involving a gamma-matrix structure $\Ga$
and derivatives $i\prt$ is coupled to 
the expectation value of a Lorentz tensor $T$,
which breaks Lorentz and CPT symmetry.
The coupling coefficient involves 
a dimensionless coupling constant $\la$
and an appropriate power of some large scale $M$,
such as the Planck or compatification scale.

\medskip\noindent \it
Is there an extension of the standard model
that includes these effects?
\rm
\\
A general extension of the standard model,
including Lorentz-breaking terms both with and
without CPT violation,
has been obtained.\cite{cksm}
The extra terms maintain the usual
SU(3) $\times$ SU(2) $\times$ U(1) gauge invariance 
and are power-counting renormalizable.
A framework has also been given 
for treating theoretically the effects
of spontaneous CPT and Lorentz breaking.

\section{Experimental Tests}

\noindent \it
How can CPT be tested experimentally to high precision?
\rm
\\
Oscillations of neutral mesons $P$, 
where $P$ is one of
$K$, $D$, $B_d$, or $B_s$,
are sensitive probes of CPT violation 
by virtue of their interferometric 
nature.\cite{kp3,ck2}$^{\!-\,}$\cite{kv}
The time evolution of the oscillations
is governed by a $2\times 2$ effective hamiltonian $\La$.
Conventional quantum mechanics 
allows in principle two complex CP-violating parameters
to appear in $\La$:
the usual CP- and T-violating parameter $\ep_P$
that preserves CPT,
and a CP- and CPT-violating parameter $\de_P$
that preserves T.
Experiments bounding the value of $\de_P$ 
can test CPT to high precision.

\medskip\noindent \it
How do the theoretical modifications
affect experimental observables?
\rm
\\
Within the CPT-violating extension of the standard model,
nonzero values of $\de_P$ emerge from small corrections
to conventional perturbative calculations.
For a given $P$ system,
it turns out that $\de_P$ 
is given by\cite{kp2,kp3}
\beq
\de_P = i 
\fr{h_{q_1} - h_{q_2}}
{\sqrt{\De m^2 + \De \ga ^2/4}}
e^{i\hat\ph}
\quad .
\label{b}
\eeq
Here,
the experimental observables
$\De m$ and $\De\ga$ are mass and rate differences,
with $\hat\ph = \tan^{-1}(2\De m/\De \ga)$.
The parameters $h_{q_j}=r_{q_j}\la_{q_j}\vev{T}$
are determined by coefficients 
of terms in the standard-model extension
and by factors $r_{q_j}$ from the quark-gluon sea.

\medskip\noindent \it
Are there definite signals from spontaneous CPT violation?
\rm
\\
Assuming hermiticity of the standard-model extension, 
the $h_{q_j}$ are real.
This implies the condition\cite{kp2,kp3}
\beq
\Im \de_P = \pm \fr{\De\ga}{2\De m} ~\Re\de_P
\quad . 
\label{c}
\eeq
Moreover,
the severity of the suppression factor for Planck-scale effects
suggests direct CPT violation in $P$-meson decay amplitudes 
is unobservable.
The relation \rf{c} for indirect CPT violation 
and the absence of direct CPT violation
are signatures for spontaneous CPT violation
in any $P$ system.

In addition,
the CPT-violating couplings 
in the standard-model extension 
seem likely to differ substantially for distinct quarks,
as do the Yukawa couplings.
The CPT-violating quantities $\de_P$ 
could therefore vary significantly for different $P$.
This means CPT should be tested in 
more than one neutral-meson system.
Given the sparsity of present bounds on CP violation
in the $B_d$ system,
it is even possible that $|\de_{B_d}| > |\ep_{B_d}|$,
in which case CPT effects would dominate conventional CP ones
in the proposed $B$ factories.

\medskip\noindent \it
What are the current limits and prospects for future tests?
\rm
\\
The kaon system offers the best CPT bound from neutral mesons.
The published limits\cite{pdg,expt1}
on $|\de_K|$ are of order $10^{-3}$.
Completed experiments
(e.g., CPLEAR at CERN),
ongoing ones
(e.g., KTeV at Fermilab),
and ones currently being designed
are likely to improve the bounds in the near future.

Mixing has not yet been seen in the $D$ system,
and dispersive effects make theoretical predictions uncertain.
In favorable conditions
some tests of CPT symmetry could be feasible,\cite{kp3,ck2}
perhaps even with current data
and probably with statistics available 
within the next decade.

The $B_d$ system might involve the largest CPT violation
because it includes the heavy $b$ quark.
Enough data to bound $\de_{B_d}$ at the level of order 10\% 
have already been obtained in the CERN LEP experiments 
and in CLEO experiments at Cornell.\cite{kp3,ck1,kv}
Indeed,
the OPAL collaboration at CERN has very recently 
placed a bound\cite{expt2}
on $\Im\de_{B_d}$ of about $2 \times 10^{-2}$.
The many $B$-dedicated experiments now being developed
are likely to improve this bound considerably.

\medskip\noindent \it
Are there any tests in systems other than neutral mesons?
\rm
\\
Several possibilities exist,
including signals that might emerge
from the CPT-violating extension of quantum electrodynamics
implied by the standard-model extension.\cite{cksm,bkr}
For example,
CPT violation can potentially be tightly constrained
by experiments establishing the difference between
the electron and positron anomalous magnetic moments.
Bounds could be placed 
on leptonic parameters for CPT violation 
that are comparable to those in neutral mesons.
Further bounds may emerge from photon properties.
It is important to consider a variety of tests
because the standard-model extension allows 
distinct parameters to control effects in the different sectors.

\medskip\noindent \it
What does CPT violation imply about the observed baryon asymmetry? 
\rm
\\
Conventional baryogenesis requires nonequilibrium processes
and C- and CP-breaking interactions.\cite{ads}
However,
an acceptable mechanism for baryogenesis in thermal equilibrium
might emerge from terms of the form \rf{a}
under suitable conditions.\cite{bckp}
A large asymmetry could be produced at grand-unification scales,
subsequently being diluted 
to the observed value through sphaleron or other effects.

\section*{Acknowledgments} 

I thank Orfeu Bertolami, Robert Bluhm, Don Colladay, 
Rob Potting, Neil Russell, Stuart Samuel, 
and Rick Van Kooten for collaborations.
This work was supported in part
by the United States Department of Energy 
under grant number DE-FG02-91ER40661.

\vfill\eject

\end{document}